\documentclass[preprint,preprintnumbers,amssymb,amsmath,aps,superscriptaddress]{revtex4}[10pt]
\usepackage{graphicx}
\usepackage{epstopdf}
\usepackage{dcolumn}
\usepackage{bm}
\begin{document} 
\pagestyle{myheadings}
\input epsf.tex
\newcommand{\beq}{\begin{eqnarray}}
\newcommand{\eeq}{\end{eqnarray}}
\newcommand{\nn}{\nonumber}
\def\ltap{\ \raise.3ex\hbox{$<$\kern-.75em\lower1ex\hbox{$\sim$}}\ }
\def\gtap{\ \raise.3ex\hbox{$>$\kern-.75em\lower1ex\hbox{$\sim$}}\ }
\def\CO{{\cal O}}
\def\CL{{\cal L}}
\def\CM{{\cal M}}
\def\tr{{\rm\ Tr}}
\def\CO{{\cal O}}
\def\CL{{\cal L}}
\def\CM{{\cal M}}
\def\mpl{M_{\rm Pl}}
\newcommand{\bel}[1]{\be\label{#1}}
\def\al{\alpha}
\def\bt{\beta}
\def\eps{\epsilon}
\def\eg{{\it e.g.}}
\def\ie{{\it i.e.}}
\def\mn{{\mu\nu}}
\newcommand{\rep}[1]{{\bf #1}}
\def\be{\begin{equation}}
\def\ee{\end{equation}}
\def\bea{\begin{eqnarray}}
\def\eea{\end{eqnarray}}
\newcommand{\eref}[1]{(\ref{#1})}
\newcommand{\Eref}[1]{Eq.~(\ref{#1})}
\newcommand{\gsim}{ \mathop{}_{\textstyle \sim}^{\textstyle >} }
\newcommand{\lsim}{ \mathop{}_{\textstyle \sim}^{\textstyle <} }
\newcommand{\vev}[1]{ \left\langle {#1} \right\rangle }
\newcommand{\bra}[1]{ \langle {#1} | }
\newcommand{\ket}[1]{ | {#1} \rangle }
\newcommand{\ev}{{\rm eV}}
\newcommand{\kms}{{\rm km/s}}
\newcommand{\kev}{{\rm keV}}
\newcommand{\Mev}{{\rm MeV}}
\newcommand{\gev}{{\rm GeV}}
\newcommand{\tev}{{\rm TeV}}
\newcommand{\mev}{{\rm MeV}}
\newcommand{\mnu}{\ensuremath{m_\nu}}
\newcommand{\mlr}{\ensuremath{m_{lr}}}
\newcommand{\acc}{\ensuremath{{\cal A}}}
\newcommand{\mav}{MaVaNs}
\newcommand{\disc}[1]{{\bf #1}}


\title{Inelastic Dark Matter in Light of DAMA/LIBRA}
\preprint{NSF-KITP-08-118}
\author{Spencer Chang}
\email{chang@physics.nyu.edu}
\affiliation{
           Center for Cosmology and Particle Physics,\\
  Dept. of Physics, New York University,\\
New York, NY 10003, USA}
\author{Graham D. Kribs}
\email{kribs@uoregon.edu}
\affiliation{Department of Physics and Institute of Theoretical Science,\\ University of Oregon, Eugene, OR 97403, USA}
\author{David Tucker-Smith}
\email{dtuckers@williams.edu}
\affiliation{Department of Physics, Williams College,  \\
Williamstown, MA 01267, USA
}%
\author{Neal Weiner}
\email{neal.weiner@nyu.edu}
\affiliation{
           Center for Cosmology and Particle Physics,\\
  Dept. of Physics, New York University,\\
New York, NY 10003, USA}
\date{\today}
\begin{abstract}

Inelastic dark matter, in which WIMP-nucleus scatterings occur through a transition to an excited WIMP state $\sim$ 100 keV above the ground state, provides a compelling explanation of the DAMA annual modulation signal.  We demonstrate that the relative sensitivities of various dark matter direct detection experiments are modified such that the DAMA annual modulation signal can be reconciled with the absence of a reported signal at CDMS-Soudan, XENON10, ZEPLIN, CRESST, and KIMS for inelastic WIMPs with masses $O(100 \;\gev)$.  We review the status of these experiments, and make predictions for upcoming ones. In particular, we note that inelastic dark matter leads to highly suppressed signals at low energy, with most events typically occurring between 20 to 45 keV (unquenched) at xenon and iodine experiments, and generally no events at low ($\sim 10$~keV) energies. Suppressing the background in this high energy region is essential to testing this scenario. The recent CRESST data suggest seven observed tungsten events, which is consistent with expectations from this model.  If the tungsten signal persists at future CRESST runs, it would provide compelling evidence for inelastic dark matter, while its absence should exclude it.

\end{abstract}
\maketitle
\section{Introduction}

The DAMA collaboration has observed an annual modulation in the nuclear recoil
rates of their experiment with a confidence level that now exceeds $8\sigma$ \cite{Bernabei:2008yi}.
This observation is consistent with weakly interacting massive particles (WIMPs)
in the halo striking the target nuclei slightly less and more often as the Earth
moves with and against the WIMP wind.
The WIMP interpretation has been sharply criticized due to its apparent inconsistency
with the null results of several other direct detection experiments, including CDMS and XENON.
In this paper, we demonstrate that inelastic dark matter \cite{TuckerSmith:2001hy} continues to offer a compelling 
explanation of both the annual modulation signal as well as the lack of a 
reported signal at CDMS, XENON, ZEPLIN, KIMS, CRESST  and other experiments.

Inelastic dark matter (iDM) \cite{TuckerSmith:2001hy} was originally proposed to reconcile the 
DAMA annual modulation observation and null 
results from CDMS. The basic model is a simple extension of the standard WIMP model. 
An inelastic WIMP has two basic properties:
\begin{itemize}
\item{In addition to the dark matter particle $\chi$, there exists an excited state $\chi^*$, 
with a mass $m_{\chi^*}-m_{\chi} = \delta \approx  \beta^2 m_\chi \sim 100$~keV 
heavier than the dark matter particle.}
\item{Elastic scatterings off of the nucleus, i.e., $\chi N \rightarrow \chi N$ are suppressed, 
compared with the inelastic scatterings $\chi N \rightarrow \chi^* N$}.
\end{itemize}
The splitting $\delta$ is comparable to the kinetic energy of a WIMP in the halo.
This causes the kinematics of the scattering process to be significantly modified compared with a WIMP that scatters elastically.  

The altered scattering kinematics leads to a fundamental difference in 
how an inelastic WIMP shows up in direct detection experiments:  only those with 
sufficient kinetic energy to upscatter into the heavier state will scatter off nuclei.  
The minimum velocity to scatter with a deposited energy $E_R$ is
\begin{eqnarray}
\beta_{\rm min} &=& \sqrt{\frac{1}{2 m_N E_R}} \left( \frac{m_N E_R}{\mu} + \delta \right),
\label{eq:betamin}
\end{eqnarray}
where $m_N$ is the mass of the target nucleus and $\mu$ is the reduced mass of the 
WIMP/target nucleus system.  This minimum velocity requirement  means that experiments tend to probe
only the higher velocity region of the WIMP halo velocity distribution.  This simple modification leads to three key  features that change the relative sensitivities of dark matter experiments. Specifically
\cite{TuckerSmith:2001hy,TuckerSmith:2004jv},
\begin{itemize}
\item{ {\it Heavier targets are favored over lighter targets} - For a given $E_R$ and $\delta$, a range of velocities in the halo will be accessible to a given dark matter experiment.  \Eref{eq:betamin} shows that the range available for  heavier target nuclei will be greater than that for lighter targets. Because the WIMP velocity distribution  is expected to fall rapidly above the peak velocity of $\sim 220\; {\rm km/s}$, targets with heavier nuclei tend to have greater sensitivity than those with lighter nuclei.}
\item{{\it Modulation of the signal is significantly enhanced} - For conventional WIMPs, the modulation is typically of the order of a few percent. For inelastic WIMPs, because one is generally sampling a higher velocity component of the WIMP velocity distribution, where numbers are changing rapidly as a function of velocity, the modulation can be much higher. In the extreme limit, it can be such that there are particles available to scatter at DAMA/LIBRA in the summer, but not in the winter, yielding a $100 \%$ modulation of the signal.
}
\item{{\it The spectrum of events is dramatically changed, suppressing or eliminating low-energy events} - The spectrum of conventional WIMPs rises exponentially at low energy. As a consequence, significant sensitivity gains can be achieved by pushing the energy threshold of the experiment to ever lower energy. With inelastic DM, however, the kinematical changes cause the signal to be suppressed or even eliminated at sufficiently low recoil energies. The result is a spectrum which can peak at 20 keVr and above.}
\end{itemize}
In this paper we demonstrate that together, these effects can reconcile DAMA/LIBRA with the null results of other experiments. The CRESST experiment, with its tungsten target, has a robust sensitivity. We shall see that the seven tungsten events seen there are consistent with expectations from this scenario.

Before proceeding to quantitative studies of the different experiments, we consider qualitatively how the above differences from conventional WIMPs reconcile the various experimental results.
As a first example, we consider the limits of CDMS (Ge) on DAMA. As germanium is 
significantly lighter than iodine, the sensitivity of CDMS is considerably reduced. Indeed, for some regions of parameter space, no WIMPs  in the halo are capable of inelastic scattering at CDMS, while some are  capable of scattering at DAMA/LIBRA. 
In practice, there is some residual sensitivity in CDMS to inelastic WIMPs 
at the high-velocity edge of the halo distribution.  This suggests that bounds
on inelastic WIMPs are rather sensitive to the velocity cutoff; we treat this carefully below.

In contrast to CDMS, the XENON experiment should be sensitive to inelastic nuclear 
recoils in a similar way to the DAMA experiment due to the close proximity of the respective 
target nucleus masses (Xe and I).  However, here there are two crucial differences between the 
two experiments' reported results:  (1)  the XENON experiment's energy threshold 
is far smaller than DAMA's (unquenched) energy threshold, and (2) XENON 
took their data between October 6 to February 14, which happens to be roughly during 
the time when the Earth travels \emph{with} the galactic WIMP wind, 
thus reducing average WIMP velocity.

Due to a combination of effects from kinematics and nuclear form factors, conventional WIMPs have a rapidly falling spectrum as a function of energy. This has led to a significant push to lower the energy threshold of experimental searches, in order to optimize sensitivity. 
XENON's bound on the cross section, for instance, is especially strong because of the dearth of events from 4.5 - 14.5 keV.  By comparison, interpreted as unchanneled iodine scatters, the unquenched energy range for the DAMA annual modulation signal
is roughly 22-66 keV.  Hypothesis-testing approaches for elastic WIMPs can 
achieve very 
stringent limits, even in the presence of background events at high energy, if the low energy region is under good control.

In contrast, an inelastic WIMP that leads to an annual modulation in DAMA
often recoils off nuclei with energies in a range that is nearly \emph{orthogonal} 
to the range used by XENON to set their bounds on elastic scattering WIMPs.  
This is particularly interesting as XENON's blind search identified nine candidate events between 
14.5 keV to 26.9 keV, and (outside their analysis range) another fourteen candidate events between
26.9 keV to 45 keV.  This spectrum is certainly inconsistent with the 
distribution of nuclear recoils expected from an 
elastic scattering WIMP.  Yet, we demonstrate that inelastic WIMPs with masses 
$O$(100 GeV) can be made consistent with the 
DAMA annual modulation, leaving no trace in the low energy bins of XENON, 
while contributing between a few to tens of nuclear recoil events in the XENON 
high energy bins.  Thus, we find that XENON does not rule out inelastic dark matter
precisely because they have twenty-three events in the signal region where inelastic WIMPs
would be expected to be found.

Finally, the heightened sensitivity to the high-velocity part of the WIMP halo velocity
distribution causes the modulation effect to be significantly enhanced.
This makes the time of year of an experiment's run especially important. If the modulation is a few percent, the precise dates of data taking are of little consequence, but in the presence of $O$(1) modulation, it can clearly change the limits. In the extreme case, where at some times of the year there are no dark matter particles capable of scattering, it can remove all sensitivity of (in this case) XENON to DAMA signals.

These qualitative arguments are borne out by more careful analyses. The layout of the paper is as follows: in section \ref{sec:indep} we discuss, without reference to halo models, the relative implications of unmodulated signals to the DAMA modulation signal, and show that consistency of the DAMA signal arising from quenched iodine scatters requires a model with low-energy events suppressed, and modulation enhanced well beyond the standard $\sim 2\%$ level expected for a standard WIMP. In sections \ref{sec:models} and \ref{sec:kine}, we discuss models for iDM and the kinematical implications of iDM at various experiments, which include such low-energy suppressions and enhanced modulation. In section \ref{sec:limits} we discuss the limits arising from various experiments, and find the allowed parameter space for these models. We find significant regions of parameter space open, with masses $\sim 50-200\; \gev$, which are otherwise closed in the absence of inelasticity. Finally, in section \ref{sec:discussion}, we consider the implications of the limits for current and future experiments, and conclude.

\section{Astrophysics-Independent Implications of DAMA at other Dark Matter Experiments}
\label{sec:indep}
Before we discuss the particular scenario of inelastic dark matter in more detail, we would like to point out some of the implications that DAMA has for other similar mass target experiments (i.e., those with iodine or xenon targets), which are essentially independent of astrophysical assumptions.  Later, in section \ref{sec:limits}, we will discuss the specific limits that pertain to the inelastic scenario.  

\subsection{DAMA/LIBRA \label{sec:damamodelindpt}}
The DAMA/LIBRA publication of the results of an exposure of 0.53 ton-years of NaI(Tl) \cite{Bernabei:2008yi}, has shown continued presence of the annual modulation signature, with a current significance of 8.2$\sigma$. In the 2-6 keV range, a modulated amplitude of $0.0131 \pm 0.0016$ cpd/kg/keV is found. The tension of DAMA/LIBRA with other experiments in the context of standard WIMPs is well known \cite{Gaitskell:2004gd}, and has prompted a number of alternative explanations, including spin-dependent interactions \cite{Savage:2004fn}, light WIMPs \cite{Gelmini:2004gm,Gondolo:2005hh}, mirror dark matter \cite{Foot:2003iv}, as well as inelastic dark matter \cite{TuckerSmith:2001hy}.

DAMA's new data also include the modulation spectrum, which is important for our analysis. This allows analyses beyond simple rate fits where instead actual tests of the spectrum can be performed for a given model. In the context of inelastic dark matter, we will see that this significantly constrains the acceptable range of parameter space from the DAMA data alone, even before appealing to limits from other experiments.

DAMA/LIBRA's has also published its unmodulated single-hit spectrum, giving important additional information beyond the modulated signal, since for a scattering WIMP these rates can be related. 

Let us assume that the DAMA rate has a functional form 
\be
S(t) = S_0 (1 + \alpha \cos (\omega t)).
\label{eq:damarate}
\ee
Thus, the modulated amplitude is $S_m = \alpha S_0$.  In this parameterization, $\alpha=1$ is total modulation, while $\alpha = 0.02$ corresponds to the standard 2\% modulation.  Using the measured modulated amplitude in the 2-4 keV bin of 0.0223 cpd/kg/keV \cite{Bernabei:2008yi}, we can predict an umodulated rate of the size $S_0 = (0.0223/\alpha)$ cpd/kg/keV.  In the same energy range, the measured single-hit unmodulated rate is about 1 cpd/kg/keV.  Since this single-hit rate does not discriminate against electron recoils aside from requiring the single hit, there should still be a sizeable background; thus a reasonable limit is that the dark matter signal not be larger than the observed rate.  This determines that $\alpha \gsim 0.02$, which is generally consistent with modulation, but requires a significant fraction of the observed rate to be arising from genuine WIMP events. Such constraints can be even more stringent if the signal is rising in the lower energy bins, which leads to a slight tension between the modulated and unmodulated rates at DAMA, assuming standard modulation.  This expectation will be borne out in the more detailed analysis of section \ref{sec:limits}, and suggests that enhanced modulation from inelasticity may provide a more natural fit with the lack of a significant rise in the DAMA unmodulated rate.

\subsection{A quasi-model independent comparison between XENON/ZEPLIN and DAMA/LIBRA iodine scatters \label{sec:modelindpt}}
Because there are always a number of assumptions that go into any analysis, it is often difficult to know how to interpret relative limits in the absence of a dedicated analysis. We argue here that, because of the kinematical similarities between Xe and I, there is a model independent exclusion that can be done. A similar analysis would be true for COUPP and KIMS which both contain iodine.

Let us begin with the assumptions: we assume that the annual modulation at DAMA is arising from quenched iodine recoils, arising from some spin-independent interaction. In this section, we make no assumptions about the halo velocity distribution.

The strong constraints from XENON arise from the lack of events in the low energy data. If the DAMA data are quenched iodine scatterings, however, the 2-6 keV DAMA events implies events in XENON's 22-66 keV nuclear recoil bins. While XENON has only one event in the acceptance region below 14.5 keV, it has nine in the range $14.5 \ \kev < E_R <27 \ \kev$ and another fourteen in the range $27 \ \kev < E_R< 45 \ \kev$. To what extent does this constrain DAMA?

If we are agnostic about the rates at energies below 22 keV, we can still make a statement that averaged over the course of the year, XENON should see a rate of $S_m/\alpha$, as defined in \Eref{eq:damarate}. The reported data taking at XENON occurred during a trough in the signal (i.e., in the winter), thus the expected signal rate present in the XENON data is simply
\be
R_{XENON} \simeq \epsilon_c A_{nr} S_m (1/\alpha-1)
\ee
where $\epsilon_c, A_{nr}$ are respectively the software cut acceptance and the nuclear recoil acceptance.  Using DAMA's measured modulated amplitude in the 2-4 keV range, the published XENON acceptances, efficiencies, and exposure \cite{Angle:2007uj} and assuming the acceptance extrapolates to $\epsilon_c \times A_{nr} = 0.3$ above 27 keV\footnote{Efficiencies and acceptances are only published up to 27 keV, but 0.3 is consistent with efficiencies and acceptances above 27 keV \cite{kaixuan}.}, we find an expected number of signal events at XENON
\be
N_{XENON}\simeq  4.10 (1/\alpha-1).
\ee
Assuming all 18 XENON events between 22 and 44 keV are signal, one finds $\alpha \simeq 0.19$.  If one instead  takes the 90\% CL upper limit for 18 observed events, while simultaneously taking the 90\% CL lower value for the DAMA modulation, one finds $\alpha \gsim 0.12$.

One can run a similar analysis for the ZEPLIN-II experiment \cite{Alner:2007ja}. There, one finds a similar number, except that ZEPLIN-II ran at a more optimal time, near the peak of the modulation. Thus, we instead have
\be
N_{ZEPLIN}\simeq 4.06 (1/\alpha+1).
\ee
Comparing this with the apparent twenty events in what is equivalent to the DAMA range, one must have 25\% modulation if they are all signal. Again, if we take the the 90\% maximum value number of events in this range, based upon these twenty, and the 90\% lower value of the DAMA modulation in the same range, one arrives at the weaker limit of $\alpha \gsim 0.08$.

Thus, any model explaining DAMA in the context of quenched iodine scatters, which is what would arise naturally for an $O$(100 GeV) WIMP, must satisfy two important constraints: it must have few events below 14.5 keV, and it must have a modulation of at least $\sim 10\%$. These two features are what make standard WIMP explanations of DAMA in conflict with XENON and ZEPLIN. However, these are precisely the features that arise naturally in an inelastic dark matter model.

\section{Models of Inelastic Dark Matter}
\label{sec:models}
Inelastic dark matter \cite{TuckerSmith:2001hy} was originally proposed to reconcile DAMA and CDMS-SUF. As described above, it involves a dark matter candidate with an excited state with a splitting $\delta = m_{\chi^*}-m_{\chi} \approx \beta^2 m_\chi$. It is assumed that elastic (i.e., $\chi N \rightarrow \chi N$, with $N$ some nuclear target) scatters are highly suppressed, while inelastic scatters ($\chi N \rightarrow \chi^* N$) are allowed.

Such features arise  in the context of $Z$-boson exchange. For a complex scalar, such as a sneutrino, the $Z$ couples off-diagonally between the real and imaginary components, which can then be split via lepton number violating terms \cite{Hall:1997ah}. Mixed sneutrinos \cite{ArkaniHamed:2000bq} are one candidate for inelastic dark matter \cite{TuckerSmith:2001hy,TuckerSmith:2004jv}\footnote{Such models also generate Majorana neutrino masses \cite{ArkaniHamed:2000kj,Borzumati:2000mc} that tend to be too large in typical parameter regions \cite{Arina:2007tm,Thomas:2007bu}. Implementing this scenario with Dirac gauginos resolves this tension \cite{Thomas:2007bu, abhishek}}. 

Alternatively, a Dirac fermion, such as a fourth-generation neutrino, is comprised of two Majorana fermions, which are degenerate by an effective $U(1)$ symmetry. The vector coupling is similarly off diagonal between them, and so splitting them with a $U(1)$ breaking term can also naturally achieve a model of inelastic dark matter \cite{TuckerSmith:2004jv}. One supersymmetric candidate would be a neutralino in approximately R-symmetric supersymmetry \cite{changinprog}.

\section{Kinematical Effects of Inelastic Dark Matter}
\label{sec:kine}
\begin{figure*}
\includegraphics{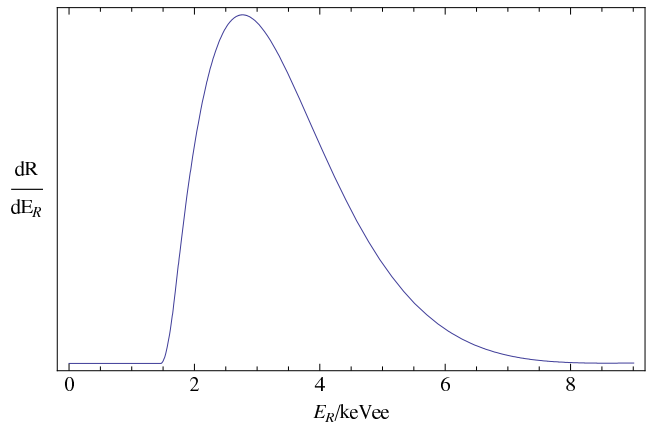}
\caption{The spectra of modulated events at DAMA for a 100 GeV WIMP with $\delta = 120$~keV. The unmodulated spectrum is similar.}
\label{fig:damaspec}
\end{figure*}

If a WIMP with mass $m_\chi$ and splitting $\delta$ scatters off of a target nucleus with mass $m_N$ and a given nuclear recoil energy $E_R$, the minimum velocity is given by \Eref{eq:betamin}.
There are two important features of this equation. First, $\beta_{min}$ is generally a falling function of $m_N$, favoring heavy targets (e.g., iodine) over lighter targets (e.g., germanium or silicon). Secondly, it has a local minimum as a function of $E_R$. That is, {\em low} values of $E_R$ can require {\em higher} values of $\beta$ in order to scatter. 
This has the result of suppressing low energy events, whereas a standard WIMP has more events at lower energies.  As an illustration, we show the spectrum of the modulated  signal at DAMA/LIBRA for a 100 GeV WIMP with $\delta = 120 \ \kev$ in figure \ref{fig:damaspec}.  The spectrum of the unmodulated signal is similar.

\begin{figure*}
\includegraphics{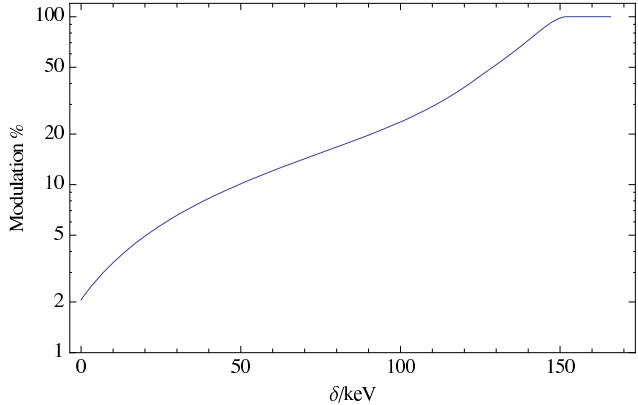}
\caption{The modulated fraction of events as a function of $\delta$ (in keV) for $m_\chi = 100$~GeV.}
\label{fig:damamod}
\end{figure*}

For the values of $\delta$ for which this spectral deformation is significant, we are dominantly sampling the high velocity tail of the velocity distribution. As a consequence, the modulation can be significantly enhanced. In figure \ref{fig:damamod}, we show the ratio of the modulated and unmodulated signals at DAMA/LIBRA in the 2-6 keV range as a function of the splitting parameter $\delta$. Remarkably, the modulation can achieve a level of 100\% at high $\delta$. This arises when there are particles whose velocity is high enough to scatter at DAMA in the summer, but not  in the winter. While this 100\% modulation is a finely tuned situation,  we will see that our fits to the DAMA signal typically require large values of $\delta$, and have significant levels of modulation, generally $\sim 30\%$. Thus, in comparing with the model-independent analysis of section \ref{sec:modelindpt}, we can already see that consistency with XENON is fairly straightforward. (For 30\% modulation, we would expect roughly half of the events in the 14.5 keV - 45 keV range to be genuine WIMP scatters.) Since we are sampling high velocity WIMPs, it is clear that a careful treatment of the escape velocity is important.

\begin{figure*}
\includegraphics[width=2.2in]{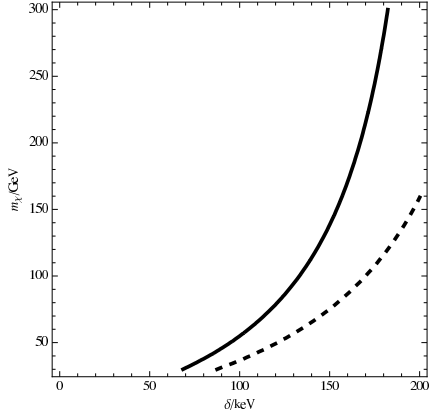}
\caption{The contours to the right of which no particles are present in the halo which can scatter at CDMS for $v_{esc}=500 \ {\rm km/s}$ (solid) and $v_{esc}= 600 \ {\rm km/s}$ (dashed), in the  $\delta -m_\chi$ plane.}
\label{fig:cdmsrange}
\end{figure*}

How we treat the velocity distribution is also important in determining the region allowed by both DAMA and CDMS.  In fact, because the limits from CDMS are so strong, consistency with DAMA usually requires higher $\delta$ as well, where there are few or no particles in the halo capable of scattering at CDMS. As an heuristic tool, we show in figure \ref{fig:cdmsrange} the values of $\delta$ and $m_\chi$ where there simply are no particles in the halo capable of scattering at CDMS for different values of the galactic escape velocity. To the right of these lines, CDMS has no sensitivity, and in the neighborhood of these lines, the CDMS sensitivity is highly suppressed. These are the principal effects that reduce the sensitivity of CDMS versus DAMA, and allow consistency between the experiments.

\section{Rates, benchmark points and spectra \label{sec:spectra}}
To calculate the rates, we employ the standard techniques for nuclear recoils \cite{Lewin:1995rx,Jungman:1995df}. The event rate at a given experiment is given by
\bea
\frac{dR}{dE_R} &=& N_T M_N\; \frac{\rho_\chi \sigma_n}{2 m_\chi \mu_{ne}^2} \frac{(f_p Z + f_n (A-Z))^2}{f_n^2}\; F^2[E_R]\; \int_{\beta_{min}}^\infty \frac{f(v)}{v}dv.
\eea

Here $M_N$ is the nucleus mass, $N_T$ is the number of target nuclei in the detector, $\rho_\chi = 0.3\; \gev/{\rm cm^3}$ is the WIMP density, $\mu_{ne}$ is the reduced mass of the WIMP-{\em nucleon} system, $F^2[E_R]$ is the nuclear form factor, and $f(v)$ is the halo velocity distribution function.  We choose to normalize our results for $f_n = f_p =1$.

While form factor dependencies at low recoil energies should be modest, the errors at large recoil energies can be significant \cite{Duda:2006uk}.  In particular, it is known that the Helm \cite{Helm} form factor can overestimate rates by $25\%$ or more. Because inelastic dark matter suppresses low energy events, keeping higher energy events, such uncertainties can be especially important.

As a consequence, we shall use the parameterized two-parameter Fermi model \cite{Duda:2006uk} for our signal (iodine), and for tungsten, which will be our most stringent constraint. For other targets, we will use a version of the Helm  form factor \cite{Lewin:1995rx,Jungman:1995df},
\begin{equation}
F^2(E_r)=\left( \frac{3 j_1(q r_0)}{q r_0} \right)^2 e^{-s^2q^2},
\end{equation}
with $q=\sqrt{2 m_N E_R}$, $s=1\ \rm fm$, $r_0=\sqrt{r^2-5s^2}$, and
$r=1.2\ A^{1/3}\ \rm fm$. 
This form factor will slightly overestimate limits from CDMS, ZEPLIN and XENON, and so is conservative. Because of uncertainties like this, it is worth remembering that precise values of exclusion plots may not be reliable. Hence, while an exclusion by an order of magnitude is likely robust, a 40\% difference from expected rates, as we shall find for CRESST, may well not be.

For the halo distribution, we assume a Maxwellian form, with a one-dimensional velocity dispersion $v_0$, truncated at $v_{esc}$. We assume that the local rotational velocity is set by the dispersion velocity, i.e., $v_{rot}=v_0$ \cite{Drukier:1986tm}. 

Previous studies of inelastic dark matter did not include physical escape velocity cutoffs. The initial analysis \cite{TuckerSmith:2001hy} included no escape velocity cutoff, while the second \cite{TuckerSmith:2004jv} made the approximation that any scattering that required $\beta_{min} c > v_{esc}$ was simply set to zero \footnote{Note that the most recent study of \cite{Petriello:2008jj} also uses this velocity cutoff prescription.}. Previously, when the DAMA spectral data were not published, and the experimental constraints were less severe, such an approximation was more reasonable. Now, however, a more rigorous treatment of the escape velocity is important, especially since simply removing forbidden events tends to overestimate rates at large $\delta$, while underestimating modulation \cite{TuckerSmith:2004jv}.

Thus, we form a velocity distribution and boost into the Earth's rest frame, using the Earth velocity given in \cite{Lewin:1995rx}, updated with the Sun's velocity in \cite{Dehnen:1997cq}. The current range for $v_{esc}$ is $498\ {\rm km/s} < v_{esc} < 608\ \rm km/s$ at $90\%$ confidence \cite{Smith:2006ym}. We will take escape velocities of $500\ \rm km/s$ and $600\ \rm km/s$, and set the velocity dispersion  to be $220\ \rm km/s$.

\subsection{Benchmark Points}
Before moving on to consider the full allowed parameter space,  in table \ref{tb:benchmark} we show a few benchmark parameter points that are consistent with all experiments.  They were obtained by taking our best fit points to DAMA's spectrum, for $\sigma_n$ and $\delta$ at a given mass, and then increasing $\delta$ until consistency with all experiments was obtained.  For all these points, $v_0 =$ 220 km/s and $v_{esc} = $ 500 km/s, where CRESST was the strongest constraint.   

\begin{table}
\begin{tabular}{|c|c|c|c|c|c|c|c|c|c|}
\hline 
\# &$m_\chi$ & $\sigma_n$ & $\delta$ & DAMA& XENON & CDMS & ZEPLIN & KIMS& CRESST\\
&&&& 2-6 keVee &  4.5-45 keV & 10-100 keV & 5-20 keVee & 3-8 keVee&12-100 keV\\
&(GeV) & ($10^{-40} \, {\rm cm^2}$) & (keV) &  ($10^{-2}$ dru) & (counts) & (counts) & (counts) & ($10^{-2}$ dru) & (counts) \\
\hline 
expt & &&& $1.31 \pm0.16$ & 24 (31.6)  & 2 (5.3) & 29 (37.2) & $5.65 \pm 3.27$ & 7 (11.8) \\ \hline 
1 &70 & $11.85 $&   119 & 0.89 &1.39 & 0 &  8.46& 0.65 & 8.76 \\
 2 & 90 & $5.75 $& 123 & 1.21 & 5.52 & 0 &  14.40& 1.52 & 9.75 \\
 3 & 120 & $3.63 $& 125 & 1.22& 9.06 & 0.13 & 18.09 & 2.18& 10.7 \\
 4 & 150& $2.92$ & 126 & 1.18 & 11.17 &0.95&  19.93& 2.53& 11.2 \\
 5 & 180& $2.67  $& 126 & 1.15 &  12.46 &1.93 & 21.01& 2.74& 11.6 \\
 6 &  250 & $2.62 $& 127 &1.11 & 14.01 & 3.60 &  23.32& 3.00 & 12.1 \\
  \hline
\end{tabular}
\caption{Parameter values and predicted experimental signals for $v_{esc}=500\; \kms$. Note that 1 dru = 1 cpd/kg/keV. In the second row, experimental observed rates and number of events are given.  These counts listed are without any background subtraction.  In parentheses are the 90\% Poisson confidence upper limits on the expected number of signal events.}
\label{tb:benchmark}
\end{table}

\begin{figure*}
\includegraphics[width=3.5in]{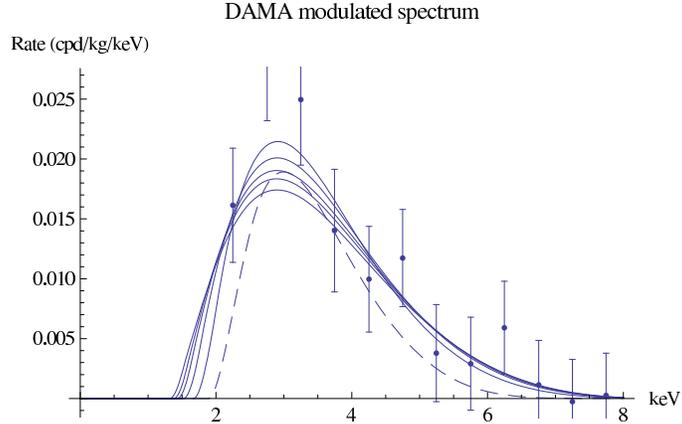} 
\caption{Modulation amplitudes for benchmark points at DAMA for $v_{esc}=500\; \kms$.  The measured modulation amplitudes of the full DAMA + DAMA/LIBRA data set are shown as data poins.  The dashed line is the 70 GeV benchmark, which is outside the 90\% confidence of the combined DAMA + DAMA/LIBRA 2-6 keV modulation, but inside the 90\% confidence region for DAMA/LIBRA 2-6 keV rates alone. For the solid lines, from the highest to lowest peak, and narrowest to broadest curves, $m_\chi = 90,120,150,180,250 \;\gev$.}
\label{fig:damamodspec}
\end{figure*}

The iodine modulated signal has overall rates consistent with that observed in the 2-6 keV range at DAMA/NaI and DAMA/LIBRA. However, quite unlike the rapidly rising rates expected over most ranges of parameter space for conventional WIMPs, the spectra turns over at low energies, see figure \ref{fig:damamodspec}. This shape is natural in iDM, and is driven by the low value of the 2-2.5 keV bin in the DAMA spectrum. Thus, even before considering other experiments, we are pushed to large values of $\delta$ where such spectral features occur.

\begin{figure*}
\includegraphics[width=3.5in]{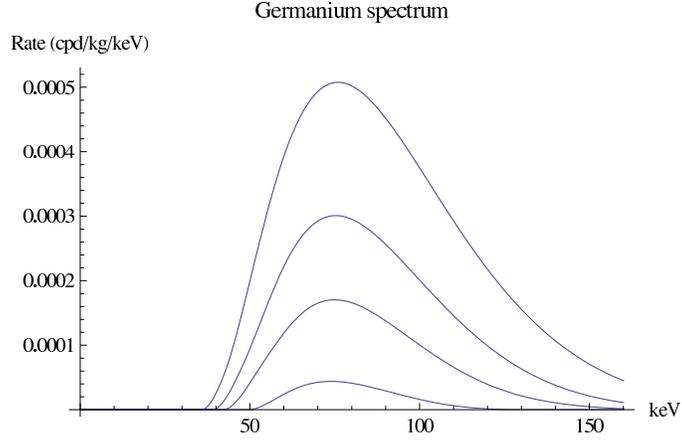}
\caption{Spectrum of events at CDMS (Ge) for $v_{esc}=500\; \kms$ (from lowest to highest signal) $m_\chi = 120,150,180,250 \; \gev$. Curves for the other benchmark points are zero. }
\label{fig:cdmsspec}
\end{figure*}

At low masses ($\sim$ 60 GeV), 
the finite galactic escape velocity can put the entire region preferred by DAMA out of the CDMS range. 
At higher masses CDMS becomes relevant, but its limits are weakened by the fact that the peak of the inelastic spectrum is located near 64 keV, where an event was observed.  This spectrum is shown in figure \ref{fig:cdmsspec}. The future ability of CDMS to test these high mass ranges will depend principally on controlling background in the 50-100 keV range.

\begin{figure*}
\includegraphics[width=3.5in]{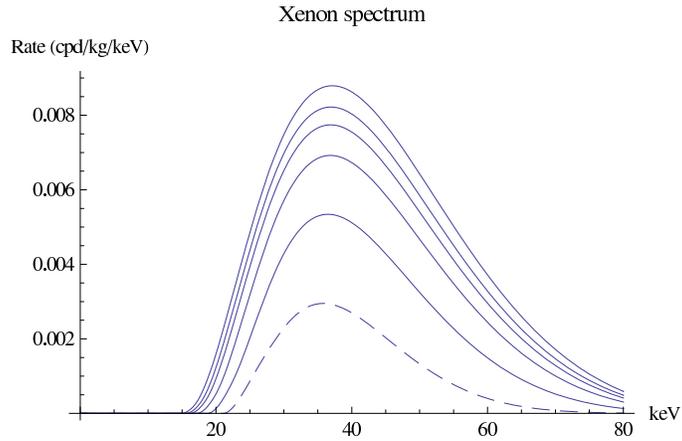}
\caption{Spectrum of events at XENON for $v_{esc}=500\; \kms$ (from lowest signal to highest signal) $m_\chi = 70,90,120,150,180,250 \; \gev$. The dashed line is the 70 GeV benchmark.}
\label{fig:xenonspec}
\end{figure*}

The XENON and ZEPLIN rates are similarly under control, and the spectrum is as required by the model independent analysis of section \ref{sec:indep}. Namely, the rates are essentially zero below 15 keV, as shown in figure \ref{fig:xenonspec}. 
As already mentioned, the DAMA spectral data also prefer this low-energy suppression.

\begin{figure*}
\includegraphics[width=3.5in]{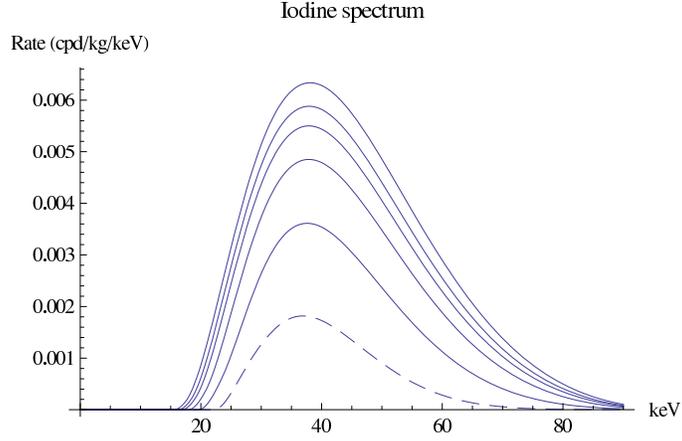}
\caption{Spectrum of events on an iodine target for $v_{esc}=500\; \kms$ (from lowest to highest signal) $m_\chi = 70,90,120,150, 180,250 \; \gev$. The dashed line is the 70 GeV benchmark.}
\label{fig:iodinespec}
\end{figure*}

KIMS rates are well below their measured value, which includes background, but often only a factor of three or so smaller. As a consequence an improvement in KIMS has a strong chance to test this scenario. We show the iodine spectrum (which is similar to the xenon spectrum) in figure \ref{fig:iodinespec}. The signal range is dominantly within the KIMS scope.

As a side note, it is often observed that the DAMA unmodulated spectrum does not rise dramatically in the region where the modulation is present. In iDM scenario, this feature is not problematic because the DAMA unmodulated spectrum is expected to go to zero at low energies, similar to figure \ref{fig:iodinespec}.

\begin{figure*}
\includegraphics[width=3.5in]{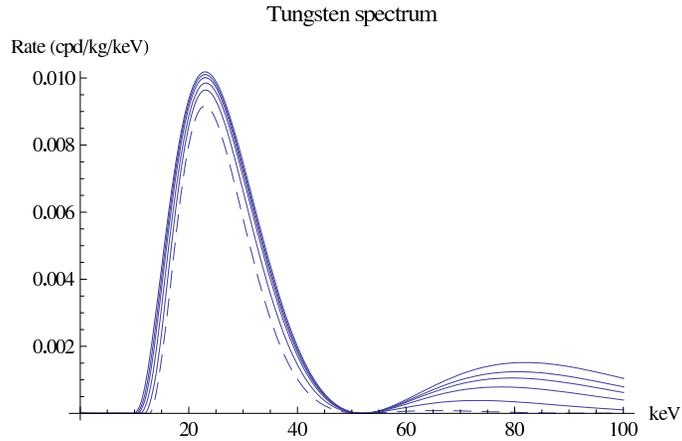}
\caption{Spectrum of events on a tungsten target for $v_{esc}=500\; \kms$ (from lowest to highest signal) $m_\chi = 70,90,120,150, 180,250 \; \gev$. The dashed line is the 70 GeV benchmark.}
\label{fig:tungstenspec}
\end{figure*}

Finally, we consider the situation at CRESST. Although the event rates are generally slightly higher than the seven events seen at CRESST, they are still within or very nearly within the 90\% Poisson confidence interval. From fig. \ref{fig:tungstenspec} we see that iDM typically produces significant event rates above 40 keV (the cutoff employed for the exclusion in \cite{Lang:2008fa}). While conventional WIMPs have 95\% of their events below 40 keV, we have for our benchmark points, 4\%, 8\%, 13\%, 16\%, 18\%, and 21\% above 40 keV for 70 GeV, 90 GeV, 120 GeV, 150 GeV, 180 GeV and 250 GeV, respectively at a tungsten experiment. Consequently, a much higher range in energies should be explored, if possible.

\section{Limits and allowed parameter space \label{sec:limits}}
In the next subsections we discuss the specific limits set by the different experiments.  Our assumptions for the data-taking periods will also be specified in the cases where the data are known to be taken non-uniformly during the year.  For CDMS, CRESST, XENON, and ZEPLIN we use the $p_{max}$ method, as described in \cite{Yellin:2002xd}, to set our limits.  This method gives comparable limits to the better known optimum interval method, but is easier to implement \cite{Yellin:2002xd}.  
 We do not take into account  finite energy resolution for any of the experiments, in particular at DAMA, where DAMA/NaI and DAMA/LIBRA may have different energy resolutions.    
As usual, all events are considered as potential signal events when determining the limit.  Finally, as a check of our methods, we have verified that our limits for $\delta=0$ are consistent with the published results from CDMS and XENON.

\subsection{DAMA}

To determine the allowed regions, we take advantage of the newly released binned data from DAMA/LIBRA \cite{Bernabei:2008yi}.  Scattering off of Na and I requires quenching factors to convert to keV electron equivalent, which we take to be $q_{Na} = .3$ and $q_{I} = .085$.  We use the twelve 0.5 keV width bins between 2.0 - 8.0 keV, and their relative uncertainties to determine a $\chi^2$ function.  We do not fit to their high-energy bins as the inelastic spectrum falls off quickly at higher energy.  In our two dimensional plots, we find the best fit point by setting values both for the unplotted variable ($\delta$ or $m_{\chi}$) and the astrophysical parameters.  For a goodness of fit, we require that the best fit point on each plot have a $\chi^2_{min} < 10$, given that we have 12 bins and 2 free parameters.  To take into account regions with a reasonable $\chi^2$, we plot contours of 90\% and 99\% confidence for two parameters, corresponding to $\chi^2 = \chi^2_{min} + 4.61$ and $\chi^2_{min}+9.21$.  In particular, our benchmark points were obtained from deviating from best fit points, all with $\chi^2_{min} <5.2$, and except for the 70 GeV benchmark, within the allowed 90\% range in $\Delta \chi^2$.  

DAMA has also released the rates for their unmodulated spectrum \cite{Bernabei:2008yi}.  The collaboration has not provided an estimate of the expected background rates in these bins; the only method to discriminate against electron recoils is the requirement that only a single detector receives a hit per event.  Since this suggests a sizeable unmodulated background, we take as a limit that the expected signal rate is not larger than the measured rate in any of the first ten bins, covering the range from roughly 0.87 keV to 3.12 keV.  As alluded to earlier in section \ref{sec:damamodelindpt}, this limit is in tension with standard modulation and thus can rule out regions with small $\delta$, suggesting that for the mass range we explore, DAMA's full data set prefers inelastic dark matter over elastic dark matter.     
Finally, we have not included channeling as discussed in \cite{Bernabei:2007hw}, as our signal is zero at low recoil energies, where this effect is important. At higher recoil energies, this amounts to a small correction to the overall rate. However, since we do not have simulations to apply it to the other experiments with target crystals such as CRESST, we opt not to apply it for consistency.   

\subsection{CDMS}
For limits from the CDMS experiment, we consider data from the three runs at CDMS-Soudan.  CDMS has given the effective germanium exposures weighted for a 60 GeV mass WIMP; these numbers  approximately apply for an inelastic WIMP, since their acceptance is roughly constant over their energy range.  In particular, the first run had 19.4 kg-day \cite{Akerib:2004fq}, the second 34 kg-day \cite{Akerib:2005kh}, and the latest 5 tower run had 121.3 kg-day \cite{Ahmed:2008eu}.  
In total, these runs saw two events, one at 10.5 keV and the other at 64 keV. 
 We have not included scattering off of silicon as this is highly suppressed when the limits from germanium scattering are applied.         

\subsection{XENON}
For the limits from XENON \cite{Angle:2007uj}, we  consider not only the low energy bins between 4.5-27 keV, as analyzed in their limits, but also the high energy region from 27-45 keV \cite{kaixuan}.  From the discussion on the model-independent limits from XENON, the latter region is potentially the most relevant for seeing any WIMP that produces the DAMA signal by scattering off of iodine. To proceed, we again assume that the acceptance at high energies is a constant at 30\% as discussed in section \ref{sec:modelindpt}.  Overall, XENON observed 24 potential signal events in the 4.5-45 keV range.  
We extract limits by taking their exposure of 316.4 kg-day and folding in their acceptances and efficiencies. 
Finally, since the XENON run was from October 2006 to February 2007, we use the average of the velocity distribution over those five months, which are near the minimum of the annual modulation.

\subsection{CRESST}
We include the published results from the CRESST-II experiment \cite{Angloher:2004tr} and also include commissioning results recently published \cite{Lang:2008fa,Angloher:2008jj}.  These new results gives the strongest constraint on inelastic dark matter.  
From their older analysis \cite{Angloher:2004tr}, we use both the DAISY and JULIA detectors which had an exposure of 20.5 kg-day and acceptance of 0.9.  Recent analyses have only included the DAISY detector, which had no events from 12-50 keV with Q=40, and only one when varying the quenching factor within experimental uncertainty, by arguing that the JULIA detector had poorer light resolution. However, it is not clear that an explicit light resolution cut was established prior to that analysis, or prior to the most recent run, and it seems most proper to include the full set of the initial run in this analysis.  From the earlier run, we use the five events observed between 12 to 50 keV, which includes the event allowed by the uncertainty in the quenching factor.  From the analysis of the commissioning run \cite{Lang:2008fa,Angloher:2008jj}, we use the exposure of 47.9 kg-day and acceptance of 0.9, and include the seven observed events from 12 keV to 100 keV.  To take into account the difference in energy regions, we do not integrate the previous run over 50 to 100 keV.     

We only implement scattering off of the tungsten (W) in their ${\rm CaWO}_4$ crystals, as calcium and oxygen scatterings should be sufficiently suppressed.  Finally, the early CRESST run was from January 31st to March 23rd in 2004, and  the new run was from March 27th to July 23rd in 2007.
We therefore average over the velocity  distributions of February and March, and April through July, respectively.

The presence of seven events in this experiment is an exciting new result, because it is consistent with the $O(10)$ events expected from iDM. However, the number is slightly smaller than would have been naively expected for iDM, and so it is worth making a few comments on some particular issues of the analysis.  First of all, we have not taken into account any energy resolution of the measured values on our limits.  This can potentially have a significant effect, since moving the observed points by 1 keV (the listed energy resolution at CRESST) can impact the limit from the $p_{max}$ technique dramatically, potentially allowing more parameter space. Such uncertainties should be included in any future interval-based analyses at CRESST with such few events. Also, since this includes commissioning (non-blind) data, and because the tungsten acceptance window has not been directly measured, it may be more prudent to consider a more conservative analysis where we include all nuclear recoil events in the limit.  With such a small number of events presently included, such effects can change limits significantly, since this changes the observed nuclear recoils from seven to nine. Finally, because at CRESST, the signal region for iDM goes over a region where the form factor goes to zero, calculating the expected events for any type of interval approach to exclusion plots has subtleties which may introduce additional uncertainties.

These subtle quantitative issues aside, we are encouraged that there are a substantial number of nuclear recoil events at a well-shielded tungsten experiment.  These observed events do not seem consistent with neutron background, since such events should dominantly appear above the tungsten acceptance region.  Should future data continue to defy background expectations, it is worth noting that a joint $\chi^2$ for DAMA and the most recent seven events at CRESST, i.e. $\chi^2_{total}=\chi^2_{DAMA} + \frac{(signal-7)^2}{signal}$, has a decent fit.  For example, for our 90 GeV benchmark point, this gives a $\chi^2_{total} = 7.1$ for eleven degrees of freedom.  This means that considering both DAMA and CRESST as signal allows points with a reasonable goodness of fit.

\subsection{KIMS}
The KIMS experiment \cite{Lee.:2007qn} has a robust target (iodine in their CsI(Tl) crystals) and energy range (3-11 keV electron equivalent) to probe the iDM explanation for DAMA.  \
KIMS has extracted their rate for nuclear recoil events, but has not parameterized the background expectation in this range.  As a simplified limit, we require that the signal rate not be larger than the measured rate plus 1.64 times its error in each of the first five bins individually, covering 3-8 keV.  To do more than this requires a background estimation which is beyond our capabilities.  However, even this naive limit can be competitive with the other experiments, so it would be interesting to see what a dedicated analysis would find.       

\subsection{ZEPLIN}
Two additional experiments employing Xe are important to discuss. ZEPLIN-I \cite{Alner:2005pa} was a xenon based pulse-shape discrimination experiment, while ZEPLIN-II \cite{Alner:2007ja} was a two-phase xenon based experiment, similar in workings to XENON10. Both experiments 
placed strong constraints on standard WIMPs. 

ZEPLIN-I, with a 3.2 kg fiducial mass, placed limits of 0.49 and 0.26 cpd in the 5-7 keVee and 7-10 keVee ranges, respectively. At the time of publication, a quenching factor of 0.22 was used, making these ranges 22.7-31.8 keVr and  31.8-45.5 keVr, both of which would have significant signals from DAMA. More recent studies of the xenon quenching factor \cite{Aprile:2005mt,Chepel:2006yv}, suggest a value closer to $0.19 \pm 0.02$. For this quenching factor, the sampled energy ranges are 26.3-36.8 keVr and 36.8-52.6 keVr. For these new ranges, the above limits translate into 0.14 cpd/kg/keVee and 0.06 cpd/kg/keVee at DAMA, which are only constraining for modulation below 15\%. Including even a $1\sigma$ fluctuation in quenching, by taking 0.17, reduces this even further. Consequently, ZEPLIN-I is not competitive with the other experiments we include, so we will not include its limits on our plots.

ZEPLIN-II, with 225 kg-day of exposure, saw 29 events in the range 13.9-55.6 keVr, roughly the same range where XENON10 saw 18 events with 316.4 kg-day.  The run was assumed to be in May and June, as suggested by \cite{zepwebpage}.  If their background analyses are correct, where they predict $28.6 \pm 4.3$ events, there is a Poisson upper limit of 10.4 WIMP signal events in the region, which places a strong constraint on the allowed parameter space, ruling out significant parts of the DAMA preferred region. However, as we are treating the events in XENON10 as signal for the purposes of setting bounds, it seems most appropriate to take the same approach to ZEPLIN-II.  Therefore we will also treat the ZEPLIN observed events as signal, which by Poisson statistics limits the expected signal to be less than 37.2 events.  The limit depends sensitively on the expected background rate.  For instance, if the background normalization is off by a factor of two, giving $14.3 \pm 2.15$ events, the Poisson limit is 23.4 events which allows almost all of the DAMA 90\% CL regions.  In the end, we apply a conservative limit using the $p_{max}$ method treating all 29 events as signal, but  if it is borne out that ZEPLIN is capable of robustly modeling their backgrounds, limits taking into account background expectations will rule out larger regions parameter space.  

\subsection{Allowed Parameter Space}

\begin{figure*}
\includegraphics[width=2.5in]{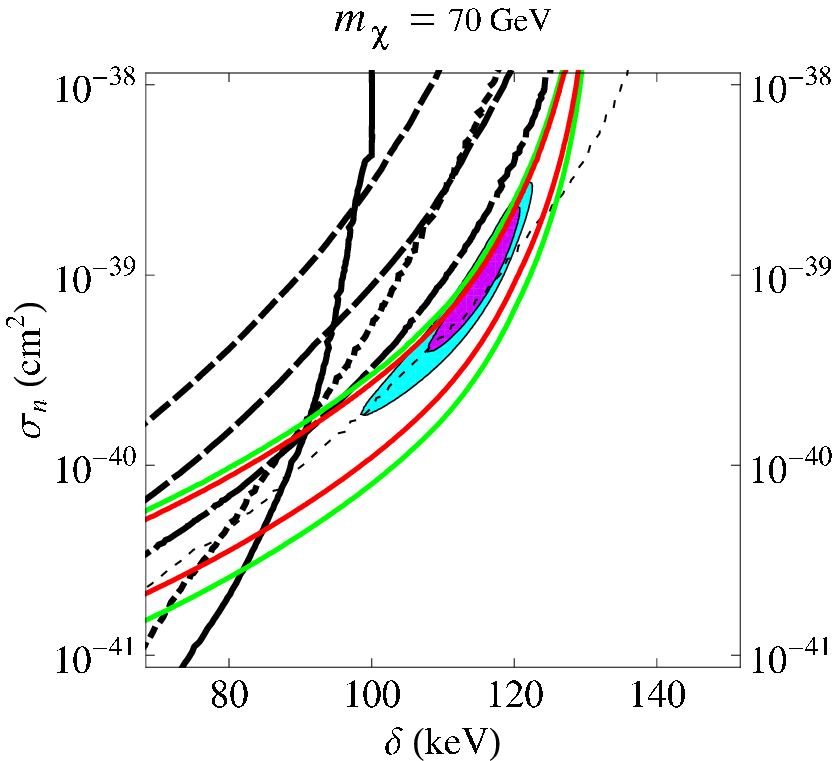} \includegraphics[width=2.5in]{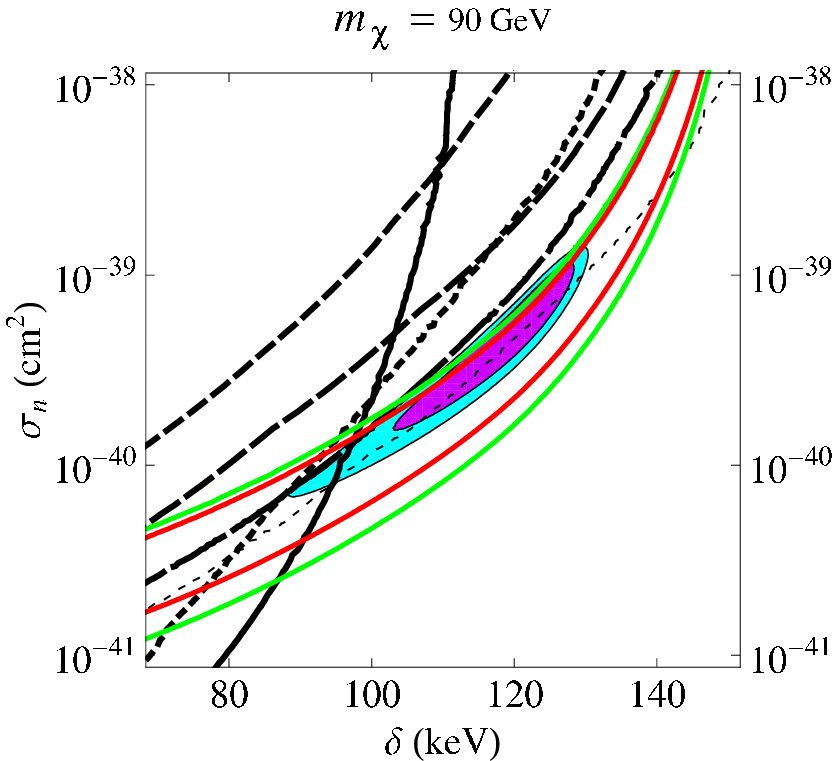}\\
\includegraphics[width=2.5in]{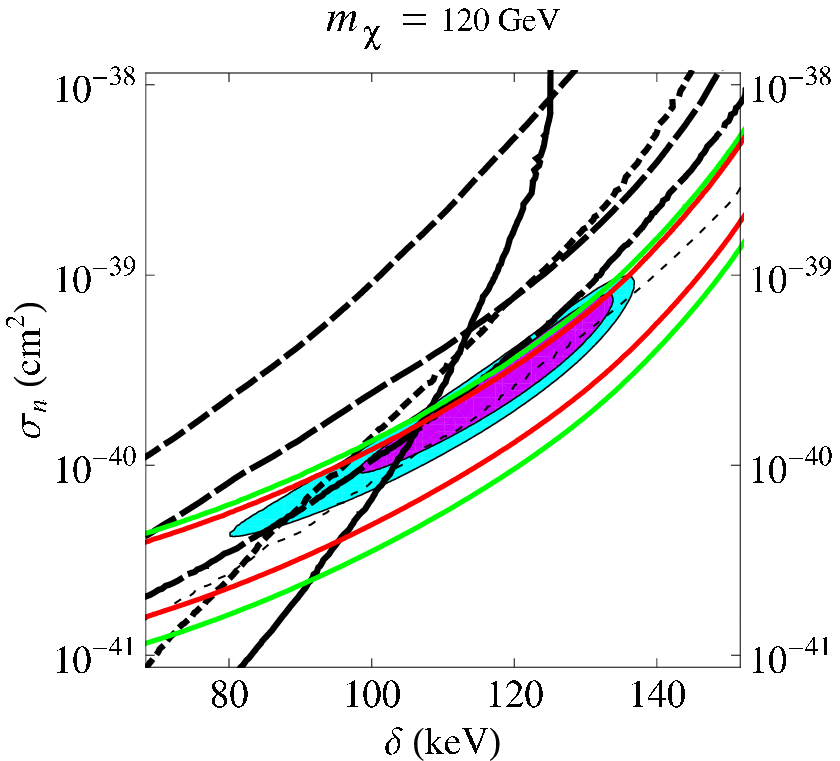} \includegraphics[width=2.5in]{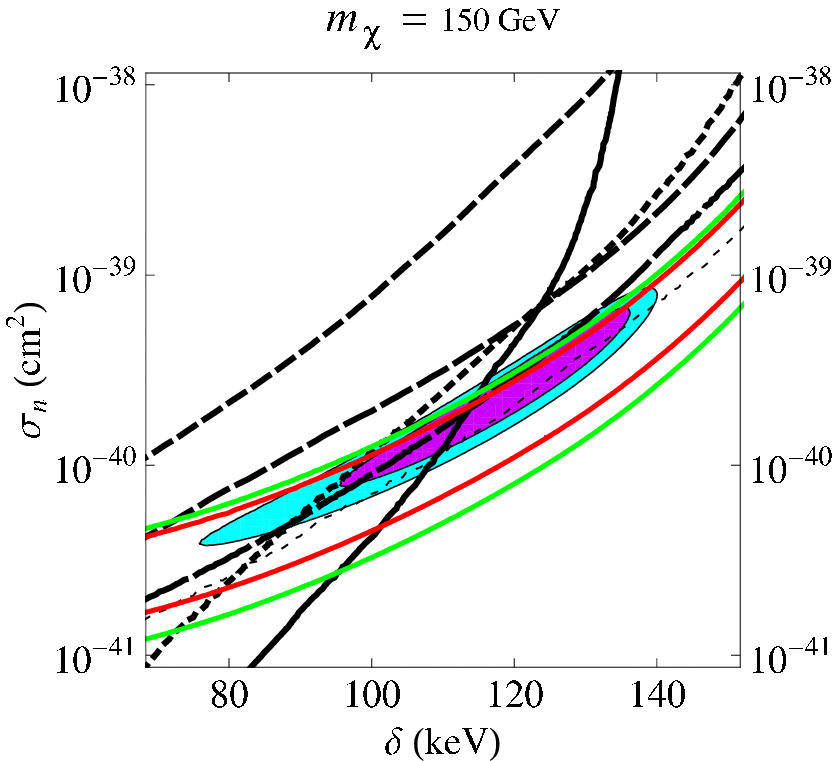}\\
\includegraphics[width=2.5in]{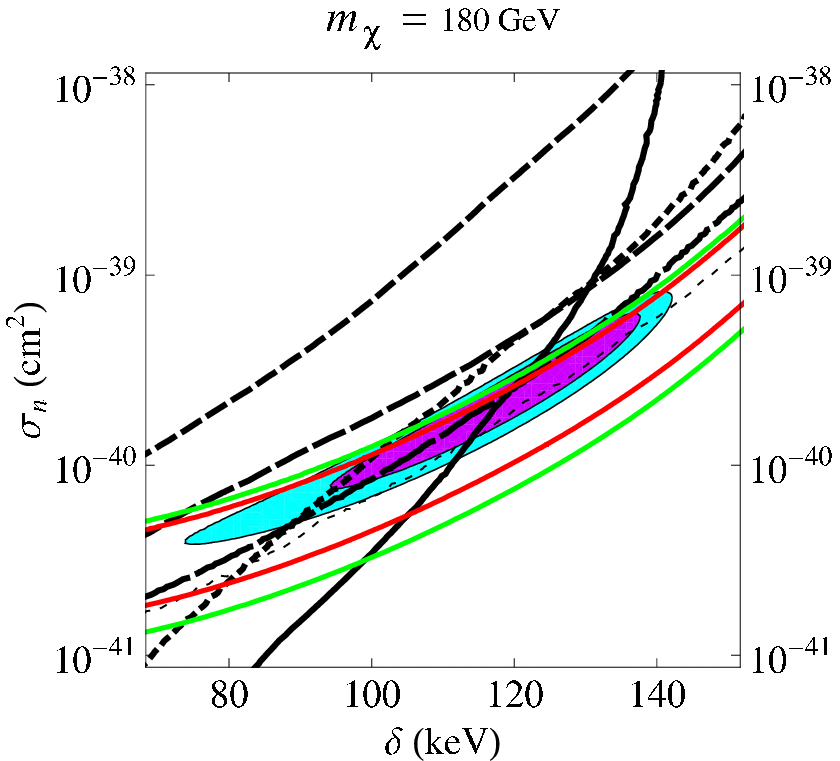} \includegraphics[width=2.5in]{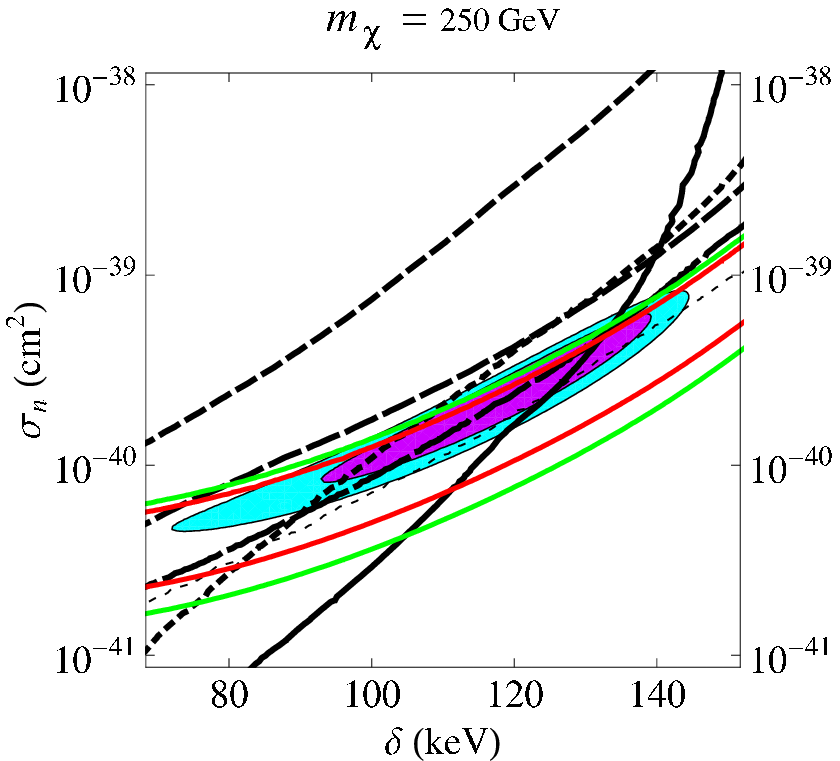}\\
 \includegraphics[width=4in]{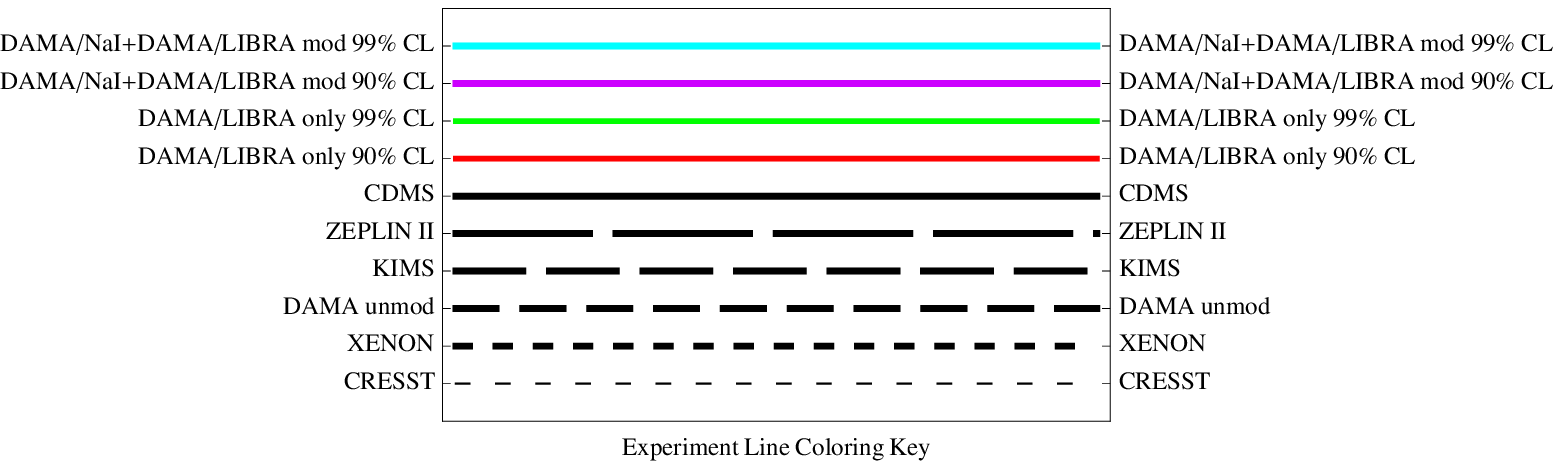}
\caption{Allowed parameter space for $v_{esc}=500 \; \kms$, $v_0=220 \; \kms$. The dark lines are from published experimental limits. The light dashed line arises from CRESST, including preliminary data from the recent commissioning run. The red and green lines are the 90\% and 99\% confidence-level contours using the DAMA-LIBRA data alone, as described in the text. }
\label{fig:param500}
\end{figure*}

Using the limits from \cite{Smith:2006ym}, figure \ref{fig:param500} shows plots with $v_{esc} =500 \; \kms$, at the low end of the 90\% confidence range for $v_{esc}$, which has the weakest constraints from CDMS.  
The given benchmark points, aside from the one with $m_\chi=70 \; \gev$ lie within the 90\% confidence interval range at DAMA, and are consistent with $p_{max}$ limits from the other experiments.  Figures with $v_{esc} =$ 600 km/s would look similar, but with a slightly smaller DAMA region consistent with all of the limits.  

\begin{figure*}
a) \includegraphics[width=3in]{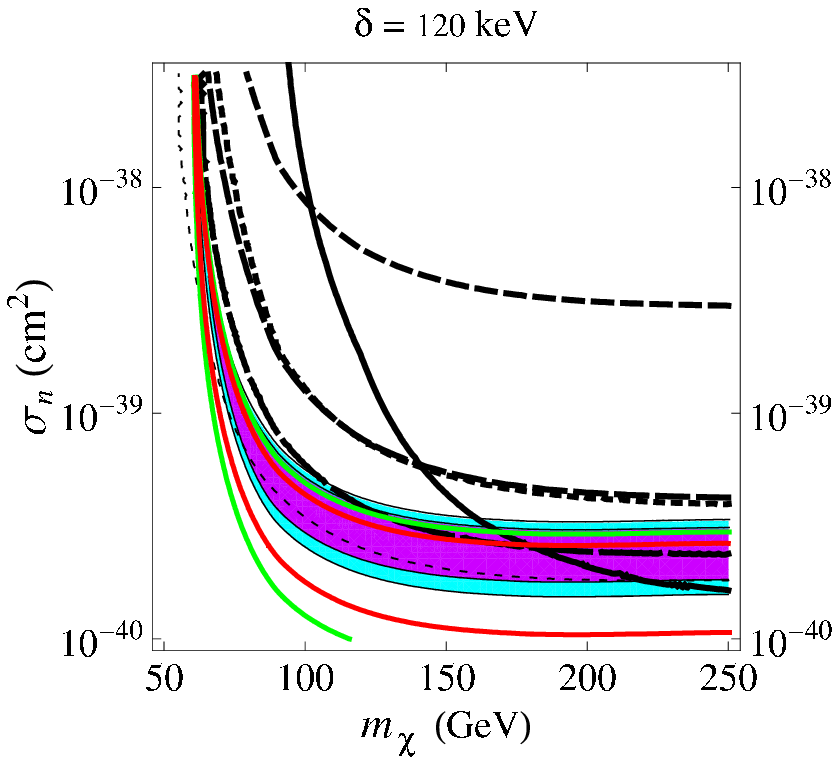} b) \includegraphics[width=3in]{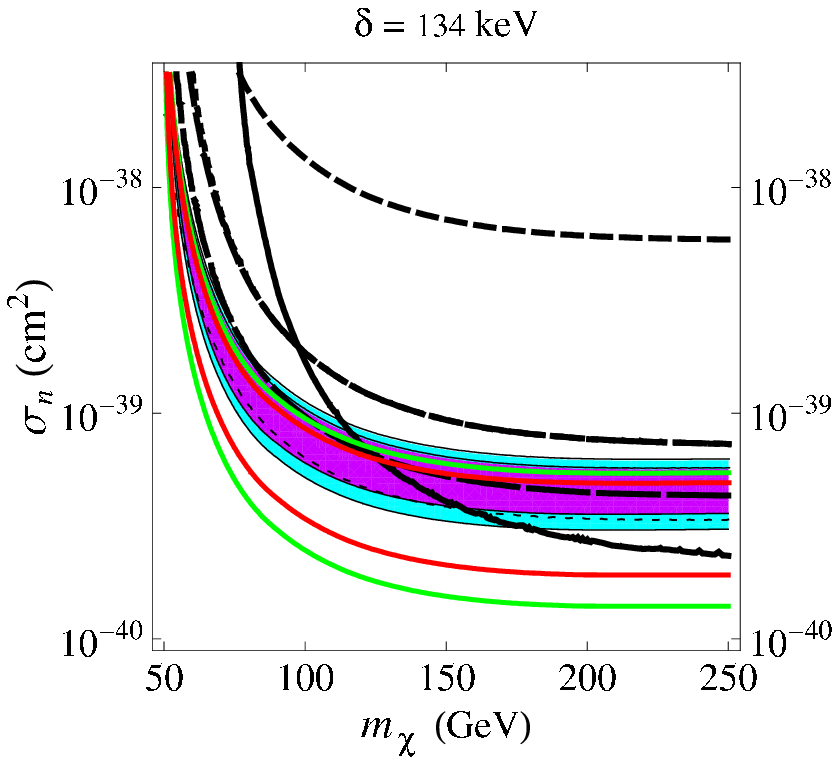}\\
\caption{Slices of allowed parameter space for the best fit values of $\delta$. a) $\delta = 120\; \kev$ and $v_{esc}=500\; \kms$. b) $\delta = 134\; \kev$ and $v_{esc}=600\; \kms$. As in figure \ref{fig:param500}, the dark lines are from published experimental limits. The light dashed line arises from CRESST, including preliminary data from the recent commissioning run. The red and green lines are the 90\% and 99\% confidence-level contours using the DAMA-LIBRA data alone, as described in the text. }
\label{fig:bfps}
\end{figure*}

Figure   \ref{fig:param500} shows that  regions of parameter space remain open. For  $v_{esc}= 500 \; \kms$ and $v_{esc}= 600 \; \kms$ we show slices of $m_\chi-\sigma_n$ parameter space in figures \ref{fig:bfps}a and \ref{fig:bfps}b at the global best fit-values of $\delta=120 \; \kev$ and $\delta=134 \; \kev$, respectively. The preliminary results from CRESST are the most constraining on the parameter space. 
The next most constraining experiments are CDMS and ZEPLIN.  
The CDMS limit is weakened by the presence of the event at 64 keV, which is near the peak expected signal at a germanium experiment (see figure \ref{fig:cdmsspec}). 
Next in strength, XENON and KIMS have similar limits. 
As stated earlier, we treat all events as signal in obtaining these limits, and stronger constraints would be obtained by taking into account the ZEPLIN background estimates.  XENON is competitive at small $\delta$, but is weaker at large $\delta$.  This is because at large $\delta$, the enhanced modulation reduces its event rate and the expected signal is peaking where XENON observed the bulk of its events.  Finally, while the DAMA unmodulated limit does not significantly constrain the inelastic scenario, it does strongly constrain the allowed regions at small $\delta$ (not shown in the plots).  This indicates that for the mass range we consider, the DAMA data alone prefer an inelastic explanation of the modulated signal over an elastic one.

For larger WIMP masses $\sim 500 -1000$ GeV, CDMS provides the dominant constraint on the inelastic scenario.  For these heavier WIMPs the CDMS constraint is still consistent with  DAMA at 99\%CL.

The annual modulation measured by DAMA/LIBRA is smaller than that measured by DAMA/NaI ($0.0107 \pm 0.0019$ versus $0.0200 \pm 0.0032$ cpd/kg/keV in the 2-6 keV recoil-energy range).  Since the discrepancy is $\sim 2 \sigma$, it is conceivable that this is simply a statistical fluctuation, and this justifies the combined analysis of the data sets. However, the DAMA/LIBRA experiment, constructed as an upgrade of DAMA/NaI experiment, could reasonably be considered as an independent measurement, and studied separately, particularly because it is under better control and monitor. Although we lack spectral data, we can compare the region favored by DAMA/LIBRA total modulation as an additional comparison. The 90\% and 99\% CL contours obtained using the annual modulation measured by DAMA/LIBRA alone, in the 2-6 and 6-14 keV bins, are shown in red and green, respectively,  in figures    \ref{fig:param500} and  \ref{fig:bfps}.   Because the annual modulation is smaller, lower cross-sections are preferred by DAMA/LIBRA and there is less tension with CRESST and other experiments. The absence of spectral information means that  $\delta$ is also less constrained by this fit.  However, assuming that the DAMA/NaI and DAMA/LIBRA spectra are similar, a full spectral analysis for DAMA/LIBRA would give a preferred region with a similar $\delta$ range as for the combined analyis, but with lower $\sigma$.


If we take into account all experimental data except the updated results from CRESST, the allowed parameter space is broader here than in the previous analysis \cite{TuckerSmith:2004jv}. This is essentially due to three reasons: (1) the DAMA parameter space has moved down slightly (due to the lower values at DAMA/LIBRA), (2)  our velocity cutoff no longer uses the unphysical approximations in \cite{TuckerSmith:2004jv}, weakening the limits, and  (3) the preliminary ZEPLIN-I limits, understood with current quenching factors, are weaker than previously thought.

It is clear from these plots that tremendous progress in the next round of experiments can be made. KIMS should have data from a 100 kg target soon, giving an exposure roughly an order of magnitude beyond what is presently analyzed. COUPP has an iodine target, and may be sensitive to this scenario. XENON10 successors, XENON100 and LUX, should be able to test this scenario if more controls are made on the higher energy backgrounds. CDMS should be able to test the higher mass ranges, even with lower galactic escape velocities. ZEPLIN-III, if backgrounds in the 20-50 keV range are smaller than ZEPLIN-II, can similarly constrain the scenario.

Most interesting at this point is the constraint from CRESST. The present curve is dominantly set not by the absence of signal, but rather by the presence of signal (specifically, the seven tungsten scattering events in the most recent CRESST run). While there is a slight tension between the size of the signal at CRESST and what is expected, without those events, CRESST would likely have excluded the iDM scenario. If continued running shows further tungsten recoils, this will be compelling evidence for iDM, while their absence should conclusively exclude it.

Three additional key points are in order: first, because of the non-standard spectrum of these theories, dedicated analyses are necessary to truly optimize the search strategies. Extrapolations of limits based upon standard WIMPs, which are dominated by the lowest energy bins, are clearly inappropriate, especially as the best fit points seem to lie at higher values of $\delta$. Secondly, since we find higher $\delta$ is consistent with the DAMA/LIBRA data, the signal itself is dominated by inelastic WIMPs at high velocities, where the halo may not be Maxwellian. As a consequence, although model-dependent limits are important, there is a need for more model independent analyses, as we have described above for XENON10, whose results depend only on particle physics, rather than astrophysical, assumptions. Finally, as we have emphasized, the compatibility of inelastic dark matter with XENON10 requires a modulation greater than roughly 10\%, and can easily be 30\% or greater. Even with their present data, both XENON10 and KIMS should be able to make tests of the very large modulation possibility, which may have been statistically impossible if it were only $\sim$ 2\%.

\section{Discussion}
\label{sec:discussion}
In light of the recent results from DAMA/LIBRA, it is important to continue to consider whether simple, viable models of dark matter exist which can reconcile the presence of the annual modulation signal with the lack of a reported signal from the many other experiments. Most recent analyses \cite{Foot:2008nw,Feng:2008dz,Petriello:2008jj,Bottino:2008mf} have focused on the light ($\sim$ few GeV) regions of parameter space, and have exploited the possibility of channeling \cite{Bernabei:2007hw} in opening low mass windows.

In contrast, the mass range for inelastic dark matter is more typical of a standard WIMP. The inelastic scattering can arise quite simply from particle physics models with approximate symmetries, and so remains an appealing possibility, quite distinct from the low-mass window.

Of particular interest is the spectral data from DAMA/LIBRA. The low ($2-2.5\; \kev$) bin shows significantly lower modulation than what usually occurs for most WIMPs, where the lowest bin often (but not exclusively) shows the highest modulation. This spectral feature arises automatically with inelastic dark matter, and drives our numerical fits to large values of $\delta$, where modulation is enhanced, and low energy rates at other experiments are highly suppressed.

We have seen that the three basic features of iDM, namely, preference for heavy targets, enhanced modulation, and suppression of low-energy events, allow the DAMA modulation to be consistent with other experiments. CDMS is suppressed due to its light target, while XENON is evaded by the suppression of events between 4.5 and 14.5 keV.

Interestingly, experiments such as ZEPLIN and KIMS, which are significantly weaker than XENON and CDMS for conventional WIMPs, are competitive in studies of inelastic dark matter. This is because they have target masses similar to iodine, and their energy range focuses on the range most relevant for studying the DAMA signal. In addition to studies similar to those presented here, studies of modulation in the presence of background events may prove quite effective, as the modulated fraction can be O(10\%) with inelastic dark matter.

Of particular importance in this vein is CRESST. Aside from a xenon-target signal shifted to a higher energy range, the second robust prediction of iDM \cite{TuckerSmith:2004jv} was the inevitable signal from tungsten scattering events. The seven events arising in the CRESST experiment are consistent with the rate expected from iDM. Should they persist in future exposure, this would be strong evidence for the inelastic nature of dark matter. Conversely, the absence of such events would exclude the inelastic explanation of the DAMA modulated signal.  .   

Ultimately, this scenario makes two clear predictions: a signal rate in the 20-50 keVr range for iodine and xenon targets, and significant rates on tungsten targets, whose signal would naturally be peaked at 20-30 keV, and extending up to possibly 80 keV. Germanium targets still have tremendous reach in their next rounds, and such a signal would be peaked at high ($\sim$ 70 keV) energies. 

The nature of dark matter remains one of the most important questions in physics. We have seen here that the inelastic dark matter scenario continues to provide an explanation of the DAMA modulation, consistent with the  results of other experiments. The robust predictions of this scenario make it exceedingly testable, and consequently, the next generation of experiments should determine if the dark matter is inelastic.

\vskip 0.15in
\noindent{{\em Note added:} As this work was being prepared, \cite{Petriello:2008jj} appeared, which also considered the inelastic scenario and concluded that the 50-200 GeV window for inelastic dark matter was excluded by CDMS and XENON. We disagree with this conclusion.  We believe the different results obtained in \cite{Petriello:2008jj} may have arisen from the approximations made there for the halo escape velocity and for the form of the modulation.} 
\vskip 0.2in
\noindent {\bf Acknowledgments}
The authors would like to thank Aaron Pierce for extensive and useful discussions. The authors would like to thank Kaixuan Ni for very insightful discussions on XENON, and providing us the information on nuclear recoils above 27 keV. The authors would further like to thank Sunkee Kim for consultation and information on the KIMS experiment. The authors would like to thank Pierluigi Belli for important information on the DAMA/LIBRA results, and Rafael Lang for discussions on CRESST. NW and GDK acknowledge the support of the KITP Santa Barbara, where some of this work was performed, and the support in part by the National Science Foundation under Grant No. PHY05-51164.  NW thanks the Aspen Center for Physics for their hospitality, where this work was completed. GDK was supported by the Department of Energy under contract DE-FG02-96ER40969. The work of DTS was supported by NSF grant 0555421.
SC and NW were supported by NSF CAREER grant PHY-0449818 and DOE grant \# DE-FG02-06ER41417.

\end{document}